\documentclass[letterpaper,11pt]{article} 

\usepackage{opticameet3} 

\newcommand\authormark[1]{\textsuperscript{#1}}

\usepackage[T1]{fontenc}
\usepackage[table,dvipsnames]{xcolor}
\usepackage[most]{tcolorbox}
\usepackage{subcaption}
\usepackage{etoolbox}
\usepackage{booktabs}
\usepackage{multirow}
\usepackage{amsmath,amsfonts,amsthm}
\usepackage{graphicx}
\usepackage{adjustbox}
\usepackage{duckuments}
\usepackage{enumitem}
\usepackage{algorithm}
\usepackage{algpseudocode}
\usepackage{hyperref}
\usepackage{tabularx}
\usepackage{minitoc}
\usepackage{multicol}
\usepackage[numbers,sort&compress]{natbib}
\usepackage{xspace}
\usepackage{array}

\newcommand{\ModelName}{\textsc{BioGen}\xspace}

\usepackage{tikz}
\usetikzlibrary{shapes.geometric,arrows.meta,positioning,calc,arrows,shapes}

\usepackage{pifont}
\makeatletter
\@ifundefined{theorem}{
  
}{}
\@ifundefined{lemma}{
  
}{}
\@ifundefined{corollary}{
  
}{}
\@ifundefined{proposition}{
  
}{}
\makeatother

\begin{document}

\maketitle

\title{\ModelName: Evidence-Grounded Multi-Agent Reasoning Framework for Transcriptomic Interpretation in Antimicrobial Resistance}

\author{
Elias Hossain,\authormark{1,*}
Mehrdad Shoeibi,\authormark{1}
Ivan Garibay,\authormark{1}
and Niloofar Yousefi\authormark{1}
}

\address{
\authormark{1}College of Engineering and Computer Science, University of Central Florida, Orlando, FL, USA
}

\email{\authormark{*}mdelias.hossain@ucf.edu}

 
\begin{abstract}
Interpreting gene clusters derived from RNA sequencing (RNA-seq) remains difficult in functional genomics, particularly in antimicrobial resistance studies where mechanistic context is needed for downstream hypothesis generation. Conventional pathway enrichment methods summarize co-expressed modules using predefined categories, but they often yield limited results and do not provide cluster-specific mechanistic explanations linked to primary literature. We present \ModelName, an evidence-grounded multi-agent framework for post hoc interpretation of RNA-seq transcriptional modules that integrates biomedical retrieval, structured interpretation, and multi-critic verification. \ModelName organizes knowledge from PubMed and UniProt into traceable cluster-level explanations with explicit evidence reporting and confidence tiering. On the primary \textit{Salmonella enterica} dataset, \ModelName achieved strong grounding and biological coherence, with BERTScore 0.689, Semantic Alignment Score 0.715, KEGG Functional Similarity 0.342, and hallucination rate 0.000 compared with 0.100 for the LLM-only baseline. Across four additional bacterial RNA-seq datasets evaluated under the same fixed pipeline configuration, \ModelName preserved zero hallucination despite variation in semantic alignment across organisms and transcriptomic settings. In a controlled multi-dataset comparison against representative open-source agentic AI baselines, \ModelName was the only framework that consistently maintained zero hallucination across all five datasets, whereas generic agentic baselines often achieved higher similarity-based scores but with substantially weaker traceability. These results indicate that retrieval access alone is insufficient to ensure reliable biological interpretation and that evidence-grounded orchestration is essential for transparent, source-traceable transcriptomic reasoning under distribution shift.
\end{abstract}
 
 
\section{Introduction}
\label{sec:introduction}

Interpreting RNA-seq gene clusters remains a central challenge in transcriptomics~\cite{kiselev2019challenges,williams2014rna,finotello2015measuring}. Clustering methods such as K-means and spectral clustering can group genes with similar expression patterns into transcriptional modules, but the downstream biological interpretation of these modules still relies primarily on enrichment-based analyses, including Gene Ontology over-representation analysis and Gene Set Enrichment Analysis (GSEA)~\cite{subramanian2005gene,zhao2023interpreting,chen2023hitchhikers}. These methods provide useful summaries at the level of predefined pathways or categories, yet they often offer limited insight when signals are weak, diffuse, or poorly represented in existing annotation resources~\cite{subramanian2005gene,zhao2023interpreting}. As a result, cluster-specific mechanistic interpretation frequently still depends on manual literature review and expert curation, which limits scalability, reproducibility, and traceable hypothesis generation~\cite{subramanian2005gene,chen2023hitchhikers}.

Large language models (LLMs) have recently emerged as powerful tools for biomedical text synthesis and knowledge-intensive reasoning. However, fluency alone does not ensure factual reliability. When used without external grounding, LLMs can generate plausible but unsupported biological statements, omit verifiable provenance, and introduce hallucinated content, including fabricated or misleading references~\cite{roustan2025clinicians}. Retrieval-augmented generation (RAG) addresses part of this limitation by conditioning generation on external evidence rather than relying solely on parametric memory~\cite{lewis2020retrieval}. Recent biomedical reviews similarly suggest that retrieval-augmented workflows can improve factuality and updateability, while also emphasizing continuing challenges in evaluation, attribution, and reliability~\cite{liu2025improving}. Even so, evidence-grounded, role-specialized agentic systems for transcriptomic cluster interpretation remain underexplored, especially in settings that require explicit critic-based verification and direct comparison with conventional enrichment pipelines.

This gap is particularly important in antimicrobial resistance research involving \textit{Salmonella enterica}, a major foodborne pathogen associated with a growing global resistance burden and substantial implications for public health, food safety, and treatment effectiveness~\cite{kumar2025antimicrobial,wang2025global}. RNA-seq can reveal genome-wide transcriptional programs related to virulence, resistance, and metabolic adaptation, but interpreting these programs still depends largely on statistical summaries followed by manual biological contextualization~\cite{chen2023hitchhikers,stevenin2024multi}. A framework that systematically links gene clusters to verifiable evidence, while also quantifying what is recovered beyond standard enrichment analysis, could therefore support more reliable and reproducible biological hypothesis generation.

To address this need, we introduce \ModelName, a role-specialized multi-agent framework for evidence-grounded interpretation of RNA-seq transcriptional modules. \ModelName formulates interpretation as a verification-constrained reasoning process involving three specialist components: (i) a \textbf{Retriever} that gathers supporting evidence from PubMed and UniProt, (ii) an \textbf{Interpreter} that produces cluster-level explanations grounded in retrieved sources, and (iii) a set of \textbf{Critic Agents} that evaluate factual grounding, semantic coherence, and adversarial consistency. \ModelName also employs a data-adaptive evidence-tiering strategy that assigns interpretations to \textit{High confidence}, \textit{Moderate confidence}, or \textit{Suggestive} categories based on the empirical distribution of consensus critic scores, thereby avoiding manually selected thresholds.

\ModelName is intended as an interpretive support layer rather than a standalone discovery engine. It does not claim novel biological discoveries or replace experimental validation. Instead, it organizes existing biomedical knowledge into transparent, cluster-specific interpretations that complement standard transcriptomic workflows and make the evidential basis of each interpretation explicit.

The main contributions of this work are as follows:
\begin{enumerate}
    \item We formulate post hoc RNA-seq cluster interpretation as an evidence-grounded, verification-aware reasoning problem and introduce \ModelName, a modular multi-agent framework that integrates retrieval, structured interpretation, and critic-based validation.
    
    \item We present a reproducible local implementation that combines per-cluster gene selection, literature and protein retrieval, data-adaptive evidence tiering, and traceable cluster-level reporting without relying on proprietary inference services.
    
    \item We evaluate \ModelName across five bacterial RNA-seq datasets using grounding- and relevance-oriented metrics, including BERTScore, Semantic Alignment Score, KEGG Functional Similarity where available, hallucination rate, and evidence-tier distributions, and analyze how retrieval and critic modules affect hallucination suppression and interpretive reliability under distribution shift.
    
    \item We compare \ModelName with standard enrichment baselines and representative open-source agentic AI frameworks under a shared local setup, showing that \ModelName consistently provides the strongest identifier-level traceability and zero-hallucination behavior across datasets, while generic agentic frameworks may achieve higher similarity-based scores but with substantially weaker grounding reliability.
\end{enumerate}

The remainder of the paper is organized as follows. Section~\ref{sec:related-work} reviews related work. Sections~\ref{sec:preliminaries} and~\ref{sec:system-architecture} introduce the \ModelName framework and its design rationale. Section~\ref{sec:experimental-setup} describes the experimental setup. Section~\ref{sec:result-analysis} presents results on the primary dataset, design sensitivity analyses, complementarity with enrichment analysis, cross-organism generalization, and comparison with representative open-source agentic AI frameworks. Finally, Sections~\ref{sec:discussion} and~\ref{sec:conclusion} discuss limitations, implications, and future directions.


\section{Related Work}
\label{sec:related-work}

Interpretation of RNA-seq gene expression data remains a longstanding challenge in transcriptomics, including in antimicrobial resistance research. Conventional transcriptomic analysis typically progresses from differential expression or clustering to pathway- or ontology-based enrichment methods such as over-representation analysis and Gene Set Enrichment Analysis (GSEA)~\cite{subramanian2005gene}. Although these methods are statistically well grounded and widely used, they summarize results at the level of predefined gene sets and may offer limited biological insight when signals are weak, diffuse, or incompletely represented in annotation resources. In addition, gene-level analysis may yield either no individually significant genes or long lists of genes without a clear unifying theme, leaving interpretation dependent on substantial expert curation.

In parallel, machine learning methods have been applied extensively to antimicrobial resistance prediction and related genomic tasks. Sequence-based language models, feature-based predictive models, and comparative genomics pipelines have contributed to resistance detection, susceptibility prediction, and evolutionary analysis~\cite{zhang2024large,you2025developing,ayoola2024predicting,wheeler2018machine}. However, these approaches are primarily designed for prediction or classification rather than for post hoc interpretation of transcriptional modules grounded in curated evidence.

More recently, large language models have been explored for biomedical annotation, clinical reasoning, and antimicrobial resistance interpretation~\cite{giske2024gpt}. Broader surveys also indicate increasing interest in retrieval-augmented and agentic frameworks that emphasize grounding, verification, and tool use~\cite{acharya2025agentic,garg2025designing,shafik2025health}. Domain-specific biomedical agent systems have likewise been proposed for tasks such as clinical decision support and multi-omics reasoning. However, some of these systems rely on proprietary models or partially closed pipelines, which complicates direct reproducibility and fair benchmarking. More importantly, LLM-based biomedical systems remain vulnerable to hallucination, incomplete source attribution, and limited traceability when explicit validation mechanisms are absent.

A central limitation in the current literature is that LLM-based interpretation is rarely evaluated directly against standard enrichment workflows. Existing studies often either replace enrichment analysis with free-form generation or use LLMs as annotation tools without quantifying what additional biological coverage they provide beyond pathway-based baselines. In this work, we address that limitation by treating the two approaches as complementary and explicitly measuring the additional cluster-level biological coverage that evidence-grounded agentic interpretation can provide relative to KEGG over-representation analysis.

Structured gene-signature resources such as MSigDB~\cite{liberzon2011molecular} further highlight the continuing importance of curated knowledge bases in transcriptomic interpretation. Although \ModelName currently retrieves evidence from PubMed and UniProt, integrating structured gene-set resources such as MSigDB represents a natural extension that could improve evidence specificity and broaden the range of recoverable cluster-level associations.

Overall, \ModelName is positioned as an interpretive layer rather than a replacement for standard transcriptomic analysis. The framework combines structured retrieval, local LLM-based synthesis, and multi-critic verification to produce transparent, evidence-traceable explanations for RNA-seq gene clusters that complement existing enrichment workflows. In addition, the present study emphasizes reproducible comparison against open-source baseline systems under a shared local setup, which is especially important in a landscape where some biomedical agent frameworks rely partly on proprietary or otherwise non-reproducible infrastructure.


\section{Preliminaries and Motivation}
\label{sec:preliminaries}

\ModelName is designed to support reliable post hoc interpretation of RNA-seq gene clusters by organizing and validating existing biomedical evidence. The framework is designed to operate on cluster outputs produced by standard transcriptomic workflows and is intended specifically for downstream biological interpretation, rather than for the development of new predictive or clustering methods.

Let $M \in \mathbb{R}^{s \times g}$ denote a normalized RNA-seq expression matrix with $s$ samples and $g$ genes. Instead of clustering samples, \ModelName operates on the transposed matrix $M^{\top} \in \mathbb{R}^{g \times s}$, where each row represents the expression profile of a gene across samples. A clustering procedure partitions the gene set into $k$ disjoint transcriptional modules $\mathcal{G} = \{G_1, \dots, G_k\}$, where each $G_i \subseteq \{1, \dots, g\}$ denotes a set of co-expressed gene identifiers. These modules serve as the primary units of interpretation.

For each cluster $G_i$, representative genes are selected according to within-cluster variance rather than global variance. Specifically, the top-$v$ genes are those with the highest variance across samples within that cluster, so that downstream interpretation is anchored in signals that are locally informative for the cluster under consideration.

The overall interpretive process can be written as
\[
F(G_i) = f_\mathcal{C}\!\left(f_\mathcal{I}\!\left(f_\mathcal{R}(G_i)\right)\right),
\]
where $f_\mathcal{R}$, $f_\mathcal{I}$, and $f_\mathcal{C}$ denote the Retriever, Interpreter, and Critic components, respectively. Intermediate outputs are stored in a shared cache to improve traceability and reduce redundant retrieval.

Given a cluster $G_i$, the Retriever $f_\mathcal{R}$ queries external biomedical resources and returns an evidence set
\[
E_i = f_\mathcal{R}(G_i),
\]
which consists of literature abstracts and protein annotations from PubMed and UniProt, each associated with explicit identifiers when available. The Interpreter $f_\mathcal{I}$ then synthesizes this evidence into a cluster-level interpretation
\[
T_i = f_\mathcal{I}(E_i),
\]
which summarizes putative biological themes while remaining constrained to the retrieved source material.

The Critic ensemble $f_\mathcal{C}$ evaluates $T_i$ with respect to factual grounding, semantic consistency, and adversarial robustness, producing an aggregated confidence score $\bar{s}_i \in [0,1]$. Rather than imposing a manually selected global threshold, \ModelName derives evidence-tier boundaries from the empirical distribution of $\{\bar{s}_i\}_{i=1}^{k}$ using the 33rd and 66th percentiles. Interpretations are then labeled \textit{High confidence}, \textit{Moderate confidence}, or \textit{Suggestive} according to their relative position within that distribution.


\section{\ModelName Framework}
\label{sec:system-architecture}

\ModelName is a modular multi-agent framework for post hoc interpretation of RNA-seq transcriptional modules. It operates on gene clusters produced by an upstream transcriptomic workflow and transforms them into structured, evidence-grounded biological explanations. The framework is built around three guiding principles: systematic use of external biomedical evidence, transparent intermediate reasoning, and critic-based verification before final reporting.

At a high level, \ModelName takes a gene cluster as input, retrieves relevant supporting evidence from biomedical resources, synthesizes that evidence into a structured cluster-level interpretation, and then evaluates the resulting explanation through multiple critics before assigning a confidence score and evidence tier. Figure~\ref{fig:framework} summarizes this workflow and illustrates the interaction among the clustering input, retrieval and interpretation modules, critic-based validation, and the final literature-grounded output.

\begin{figure}[h]
    \centering
    \includegraphics[width=0.80\linewidth]{biogen_framework.pdf}
    \vspace{0.8em}
    \caption{Overview of the \ModelName framework for interpretable RNA-seq analysis. Gene-level clusters are enriched with PubMed and UniProt evidence, interpreted by the InterpreterAgent using a local open-source LLM, and evaluated through a panel of Critic Agents, producing transparent, literature-grounded outputs with data-adaptive evidence-tier labels.}
    \label{fig:framework}
\end{figure}

\subsection{Core Components}
\label{sec:framework-components}

\ModelName consists of six functional components: an input layer, a retrieval layer, an interpretation layer, a verification layer, an evidence-tier assignment module, and an output layer. A lightweight memory module supports traceability and reuse of intermediate results throughout the pipeline. For presentation clarity, these components are grouped into two broader stages: input and evidence acquisition, followed by interpretation, verification, and output generation.

\subsubsection{Input and Evidence Acquisition}
\label{sec:input-evidence-acquisition}

The framework accepts a set of gene clusters $\{G_1, \dots, G_m\}$ derived from gene-level clustering of the expression matrix. No assumptions are made about the specific upstream clustering method. This design decouples interpretive reasoning from statistical partitioning and allows \ModelName to remain compatible with a range of transcriptomic workflows. Each cluster is treated as a fixed unit of downstream interpretation.

For each cluster $G_i$, the \textit{Retriever Agent} issues structured queries to PubMed and UniProt using representative gene identifiers from that cluster. The purpose of this step is to assemble an evidence pool containing literature abstracts, protein annotations, and associated database identifiers. Retrieved materials are stored in structured form and cached for reuse. The retrieval layer functions strictly as an evidence collection module and does not itself generate biological claims. The contribution of retrieval-source selection under this design is examined empirically in the design-sensitivity analyses presented in Section~\ref{subsec:design-justification}.

\subsubsection{Interpretation, Verification, and Output}
\label{sec:interpretation-verification-output}

The \textit{Interpreter Agent} synthesizes the retrieved evidence into a cluster-level textual interpretation. The output is structured to summarize functional themes, putative pathways, regulatory context, and explicit limitations. Interpretations are constrained to the retrieved sources and are not intended to introduce unsupported biological claims. A local open-source LLM, \texttt{Mistral-7B-Instruct-v0.2} with 4-bit NF4 quantization, serves as the generation backbone in order to reduce dependence on external services and improve reproducibility.

Candidate interpretations are then evaluated by three \textit{Critic Agents}, each of which examines the output from a different perspective. The \textit{Evidence-Strict Critic} performs rule-based factual validation against retrieved identifiers, the \textit{Semantic Critic} uses embedding-based similarity to assess alignment with the evidence pool, and the \textit{Adversarial Critic} applies counterfactual probing to expose potentially unsupported claims. These critics do not rewrite the interpretation. Instead, they produce scores and reliability judgments that are later aggregated into a consensus assessment.

After all clusters have been evaluated, consensus confidence scores are pooled and converted into evidence tiers using data-adaptive thresholds. Specifically, the 33rd and 66th percentiles of the score distribution define the boundaries between \textit{Suggestive}, \textit{Moderate confidence}, and \textit{High confidence} outputs. This percentile-based strategy avoids reliance on a fixed threshold that may not transfer well across organisms or experimental conditions. The empirical stability of this percentile-based tiering scheme, together with additional analyses of retrieval-source behavior and evidence specificity, is examined later in Section~\ref{subsec:design-justification}.

A structured cache stores retrieved evidence, intermediate interpretations, and critic evaluations. This memory module supports traceability, reduces redundant retrieval, and preserves intermediate results for later inspection. The final output for each cluster consists of a structured biological interpretation, explicit PubMed and UniProt references when available, a consensus confidence score, and an evidence-tier label. These outputs are intended to support biological contextualization and hypothesis prioritization rather than replace experimental validation.

\subsection{Design Rationale}
\label{sec:design-rationale}

The ordering of \ModelName's components was chosen to preserve traceability between retrieved evidence and the final interpretation. Retrieval is performed before generation so that the \textit{Interpreter Agent} operates on an explicit evidence pool rather than relying on parametric memory alone. This design reduces dependence on unsupported internal associations and ensures that candidate interpretations can be linked back to identifiable external sources.

The critic ensemble is applied after interpretation rather than before generation because the critics are intended to evaluate the complete cluster-level explanation, including its factual grounding, semantic consistency, and reporting quality. In this role, the critics assess not only whether evidence has been retrieved, but also whether that evidence has been synthesized into a coherent and sufficiently supported biological interpretation. Applying critics only at the retrieval stage would not capture this explanation-level behavior. At the same time, the current design does not preclude alternative orderings. Earlier verification of retrieved evidence, integration of structured signature databases such as MSigDB, or iterative critic-in-the-loop generation may all be useful extensions for future work. The empirical implications of these design choices are examined later in Section~\ref{subsec:design-justification}.

\subsection{Reliability Aggregation}
\label{sec:reliability-workflow}

For a given gene cluster, multiple Critic Agents independently evaluate factual correctness, semantic consistency, and adversarial robustness. Let $s_{ij}$ and $r_{ij}$ denote the confidence score and binary reliability assignment produced by critic $j$ for interpretation $i$. The aggregated confidence score is defined as
\[
\bar{s}_i = \frac{1}{m} \sum_{j=1}^{m} s_{ij},
\]
and the final reliability indicator $\bar{r}_i$ is determined by majority voting over the critic-level binary assignments.

Evidence tiers are assigned according to the score distribution across the full run:
\[
\text{Tier}(i) = \begin{cases}
\textit{High confidence} & \text{if } \bar{s}_i \geq p_{66}, \\
\textit{Moderate confidence} & \text{if } p_{33} \leq \bar{s}_i < p_{66}, \\
\textit{Suggestive} & \text{if } \bar{s}_i < p_{33},
\end{cases}
\]
where $p_{33}$ and $p_{66}$ denote the 33rd and 66th percentiles of $\{\bar{s}_i\}_{i=1}^{k}$.

\subsection{Algorithm}
\label{sec:algorithm}

Algorithm~\ref{alg:biogen_pipeline} summarizes the end-to-end workflow of \ModelName. The pipeline begins by initializing the cache, runtime environment, and agent instances. Gene-level clustering is then performed on the transposed expression matrix, and representative genes for each cluster are selected according to within-cluster variance. Evidence is retrieved for each representative gene, with cached results reused whenever available to reduce redundant queries. The Interpreter Agent then produces a structured cluster-level interpretation, which is subsequently evaluated by the Critic ensemble. After all clusters have been processed, evidence-tier thresholds are derived from the full score distribution, and the final labeled outputs are saved.

\begin{algorithm}[t]
\caption{\ModelName: Complete Agentic Pipeline}
\label{alg:biogen_pipeline}
\begin{algorithmic}[1]
\Require RNA-seq matrix $M$, number of clusters $k$, max evidence per gene $r$
\Ensure Cluster interpretations $\mathcal{I}$
\State Initialize cache $\mathcal{C}$, runtime environment, and retrieval clients
\State Instantiate Cluster, Retriever, Interpreter, and Critic agents
\State $(M', y) \gets \mathcal{A}_{\text{cluster}}.\texttt{run}(M^{\top})$ \Comment{Gene-level clustering}
\State $\mathcal{S} \gets \{\}$ \Comment{Collect scores for tier derivation}
\For{$i \gets 1$ to $k$}
    \State $G_i \gets$ top-$v$ genes by within-cluster variance
    \ForAll{$g \in G_i$}
        \If{$g \in \mathcal{C}$} Retrieve from cache
        \Else\ Retrieve and cache evidence (up to $r$ items)
        \EndIf
    \EndFor
    \State Construct structured prompt $P_i$
    \State $(T_i, c_i, r_i) \gets \mathcal{A}_{\text{interp}}.\texttt{generate}(P_i)$
    \ForAll{critic $j$ in critic ensemble}
        \State Evaluate $T_i$; store $(s_{ij}, r_{ij})$
    \EndFor
    \State $\bar{s}_i \gets \frac{1}{m}\sum_j s_{ij}$;\ $\bar{r}_i \gets \text{majority}(\{r_{ij}\})$
    \State $\mathcal{S} \gets \mathcal{S} \cup \{\bar{s}_i\}$
\EndFor
\State $p_{33}, p_{66} \gets \text{percentile}(\mathcal{S}, 33),\ \text{percentile}(\mathcal{S}, 66)$
\For{$i \gets 1$ to $k$}
    \State Assign evidence tier from $\bar{s}_i$, $p_{33}$, and $p_{66}$
    \State Post-process $T_i$ and save to $\mathcal{I}$
\EndFor
\State \Return $\mathcal{I}$
\end{algorithmic}
\end{algorithm}

\begingroup
\sloppy


\section{Experimental Setup}
\label{sec:experimental-setup}

The experimental evaluation was designed to assess five aspects of \ModelName: evidence grounding on the primary \textit{Salmonella enterica} dataset, sensitivity to key architectural design choices, complementarity with conventional enrichment analysis, generalization across additional bacterial RNA-seq datasets, and behavior relative to representative open-source agentic AI frameworks. The overall workflow of the framework is described in Section~\ref{sec:algorithm} and summarized in Algorithm~\ref{alg:biogen_pipeline}. This section presents the data sources and preprocessing procedures (Sections~\ref{sec:dataset-ethics} and~\ref{sec:preprocessing}), the evaluation metrics (Section~\ref{sec:metrics}), the KEGG over-representation analysis baseline used for comparison (Section~\ref{sec:gsea-baseline}), the computational environment (Section~\ref{sec:compute}), and the controlled settings used in the agentic baseline comparison (Section~\ref{sec:framework-configuration}). Additional implementation details are provided in the appendix, including model configurations and comparative systems (Appendix~\ref{app:model-configurations}), reproducibility and environment details (Appendix~\ref{app:reproducibility}), hyperparameter settings (Appendix~\ref{app:hyperparameters}), system and agent implementation details (Appendix~\ref{app:agent-implementation}), prompt templates (Appendix~\ref{app:prompt-design}), and an example prompt-output pair (Appendix~\ref{app:prompt-example}). Together, these sections define the shared experimental conditions underlying the primary-dataset evaluation, design-sensitivity analyses, cross-organism analysis, and agentic framework comparison.

\subsection{Data Sources}
\label{sec:dataset-ethics}

The primary RNA-seq dataset used in this study was obtained from the European Nucleotide Archive under accession \texttt{PRJEB67574} and originates from the multi-omics investigation of St{\'e}venin et al.~\cite{stevenin2024multi}. It contains transcriptomic profiles from 58 \textit{Salmonella enterica} isolates and provides a biologically well-characterized setting for evaluating cluster-level interpretation.

To assess the robustness and generalizability of the framework beyond the primary cohort, we additionally incorporated four publicly available bacterial RNA-seq datasets from the Gene Expression Omnibus (GEO). These external datasets were selected according to four criteria: relevance to antimicrobial resistance or bacterial stress-response biology, public availability of gene-level count data, sufficient sample size for cluster-based analysis, and compatibility of gene identifiers with downstream functional annotation resources.

The first external dataset, \texttt{GSE251671}, is a large-scale transcriptional compendium for \textit{Pseudomonas aeruginosa} UCBPP-PA14 generated under antibiotic perturbation and CRISPRi target depletion~\cite{romano2024perturbation}. It contains 1,059 samples and provides processed gene-level count data suitable for direct downstream analysis. The second dataset, \texttt{GSE144604}, was generated from \textit{Escherichia coli} MG1655 and profiles transcriptional responses to combinations of antibiotics and biocides, thereby capturing patterns of cross-protection and cross-vulnerability relevant to antimicrobial stress~\cite{wang2020accelerated}. This dataset contains 135 samples. The third dataset, \texttt{GSE224463}, is an \textit{E.~coli} K-12 time-course study examining wild-type and $\Delta rpoS$ responses across multiple stress conditions, including stationary-phase transition, osmotic stress, and low temperature~\cite{adams2023timing}. It contains 192 samples. The fourth dataset, \texttt{GSE55197}, is a \textit{P.~aeruginosa} PA14 transcriptomic compendium spanning 14 environmental conditions, including biofilm growth, anaerobic growth, osmotic variation, and nutrient limitation~\cite{dotsch2015pseudomonas}, which contains 47 samples.

Together, these datasets introduce variation in organism, experimental design, and transcriptomic context while remaining relevant to the broader problem of bacterial functional interpretation.

\subsection{Preprocessing}
\label{sec:preprocessing}

For the primary \textit{Salmonella} dataset, raw sequencing reads were processed using a standard RNA-seq workflow. FastQC was used for initial quality assessment, fastp for adapter trimming and filtering, BWA-MEM for alignment to the \textit{S.~enterica} Typhimurium LT2 reference genome (RefSeq \texttt{GCF\_000006945.2}), samtools for alignment processing, and featureCounts for gene-level quantification. This procedure produced a final expression matrix containing 58 samples and 4,679 genes.

In contrast, the four external validation datasets were obtained in processed count form and therefore did not require re-alignment from raw FASTQ reads. For each dataset, the released count tables were converted into a standardized matrix representation with genes as rows, samples as columns, and a unified \texttt{gene\_id} field. When the deposited processed data consisted of one count table per sample, these files were merged by shared gene identifier after removal of non-biological summary rows. Only datasets satisfying the minimum criteria of at least 20 samples and approximately 2,000 or more genes in the final matrix were retained for downstream validation. Full download and harmonization scripts are provided in the project repository.

To identify transcriptionally coherent modules, clustering was performed on the transposed expression matrix so that genes, rather than samples, served as the primary clustering units. This design aligns with the biological objective of identifying co-expressed genes that may share regulatory or functional relationships. K-means clustering with $k=10$ was applied, with the number of clusters selected using the elbow criterion based on within-cluster inertia. Representative genes for each cluster were selected according to within-cluster variance rather than global variance, thereby ensuring that each cluster interpretation was driven by its most informative local signals. The resulting clusters were treated as fixed inputs to \ModelName throughout all experiments.

\subsection{Evaluation Metrics}
\label{sec:metrics}

Evaluation was conducted using complementary metrics drawn from the natural language processing and biomedical literature. These measures assess semantic correspondence, biological relevance, factual traceability, and interpretive reliability from different perspectives. In the final unified multi-dataset analysis, we emphasize metrics that can be computed consistently across datasets and framework settings, namely BERTScore, Semantic Alignment Score (SAS), KEGG Functional Similarity where available, hallucination rate, and evidence-tier distributions.

\subsubsection{BERTScore}
\label{sec:bertscore}

BERTScore~\cite{zhang2019bertscore} was used to quantify the semantic overlap between generated interpretations and retrieved evidence through contextual token embeddings. We report the F1 variant computed with the \texttt{distilbert-base-uncased} backbone:
\[
\mathrm{BERTScore} = F_1(\mathbf{v}_{\mathrm{interpretation}},\, \mathbf{v}_{\mathrm{evidence}}),
\]
where $\mathbf{v}$ denotes token-level contextual embeddings. Higher values indicate stronger semantic overlap between the generated interpretation and the available evidence context.

\subsubsection{Semantic Alignment Score}
\label{sec:semantic-alignment}

The Semantic Alignment Score (SAS) measures cosine similarity between Sentence-BERT embeddings of the generated interpretation and the retrieved evidence pool:
\[
\mathrm{SAS} = \cos(\mathbf{v}_{\mathrm{evidence}},\, \mathbf{v}_{\mathrm{interpretation}}),
\]
where embeddings are computed using the \texttt{all-MiniLM-L6-v2} model. This metric reflects overall semantic proximity between the interpretation and the evidence. Because evidence density and specificity vary across organisms and datasets, SAS is most informative as a relative alignment signal within a given dataset and should be interpreted cautiously in cross-dataset settings.

\subsubsection{KEGG Functional Similarity}
\label{sec:kegg-func-sim}

To assess biological relevance, we compute the semantic similarity between each cluster interpretation and the KEGG functional descriptions associated with the representative genes in that cluster. Gene-function descriptions are retrieved through the KEGG REST API using organism-specific KEGG codes where available, and similarity is computed with Sentence-BERT embeddings. This metric serves as an external proxy for biological plausibility by comparing generated interpretations against curated functional annotations. Because KEGG annotation coverage varies across organisms and datasets, this metric is reported only where reliable mapping is available.

\subsubsection{Hallucination Rate}
\label{sec:hallucination-rate}

Hallucination rate measures the proportion of generated interpretations that do not contain explicitly verifiable citation support:
\[
H = 1 - \frac{|R_{\mathrm{verified}}|}{|R_{\mathrm{total}}|},
\]
where $R_{\mathrm{verified}}$ denotes interpretations containing at least one verifiable identifier in explicit citation form, such as a PMID, DOI, or UniProt accession, and $R_{\mathrm{total}}$ denotes the total number of interpretations. Lower values indicate better factual grounding. To avoid false positives from generic numeric strings or mention of database names without valid identifiers, verification required explicit identifier-formatted matches rather than loose keyword detection. In the final multi-dataset analysis, hallucination rate is the primary robustness metric because it is available consistently across datasets and directly reflects source-traceable grounding.

\subsubsection{Evidence Tier Assignment}
\label{sec:evidence-tier}

In addition to continuous metrics, \ModelName assigns each interpretation to one of three evidence tiers derived from the empirical distribution of consensus critic scores. Thresholds are defined at the 33rd and 66th percentiles of all cluster scores, producing three data-driven categories: \textit{High confidence}, \textit{Moderate confidence}, and \textit{Suggestive}. This tiering scheme provides an interpretable summary of evidential support without relying on manually selected cutoffs. Because the thresholds are derived independently within each dataset run, evidence-tier counts should be interpreted as relative within-run summaries rather than absolute cross-dataset confidence measures.

\subsection{Framework Configurations}
\label{sec:framework-configuration}

This subsection describes the controlled settings used for the agentic framework comparison reported later in the Results section. In the final comparison, all five evaluated systems, including \ModelName, were tested across the same five RNA-seq datasets: the primary \textit{Salmonella enterica} cohort (PRJEB67574) and four external bacterial datasets (GSE251671, GSE144604, GSE224463, and GSE55197). For every framework and dataset, we used the same LLM backbone (Mistral-7B-Instruct-v0.2, 4-bit NF4 quantization), the same evidence sources (PubMed and UniProt), and the same cluster inputs. Agentic orchestration was therefore the main experimental variable across systems.

To maintain comparability, all baseline systems used the same shared retrieval utilities and the same evidence pool construction policy as \ModelName. Consequently, differences in output quality are attributable primarily to framework behavior and orchestration strategy rather than differences in evidence access or backbone model selection.

The four open-source baseline systems were configured as follows. \textbf{LangChain ReAct}~\cite{annam2025langchain,yao2022react} used a single ReAct-style reasoning loop over the shared evidence context. \textbf{CrewAI}~\cite{shaikh2025llm,duan2024exploration} used a sequential two-agent design consisting of a Retriever-style evidence consumer and an Interpreter-style reasoning agent. \textbf{AutoGen}~\cite{wu2024autogen,dibia2024autogen} used a two-turn orchestration pattern between an Orchestrator agent and a BioAssistant agent with the same retrieved evidence injected into context. \textbf{Smolagents}~\cite{smolagents} used a lightweight tool-calling reasoning loop over the shared evidence inputs.

We also considered comparison against domain-specific biomedical agent systems reported in the literature. However, some of these systems depend on proprietary APIs or partially closed inference pipelines, which makes strict reproduction under the same local and controlled environment difficult. For this reason, the present comparison focuses on open-source frameworks that could be evaluated under a shared and reproducible setup. Extending the benchmark to biomedical agent systems with comparable public implementations remains an important direction for future work.

In addition to the agentic framework comparison, controlled variants of retrieval-source selection and evidence-tier thresholding were used in the design-sensitivity analyses reported in Section~\ref{subsec:design-justification}.

\section{Results}
\label{sec:result-analysis}

This section presents the empirical evaluation of \ModelName across five complementary dimensions. We first examine grounding and reliability on the primary \textit{Salmonella enterica} dataset through system ablations, evidence-grounding metrics, critic analysis, and qualitative output comparison. We then present targeted design-sensitivity analyses for key architectural choices, assess complementarity with standard enrichment analyses, evaluate generalization across additional bacterial RNA-seq datasets, and finally compare \ModelName with representative open-source agentic AI frameworks under a controlled setting.

\subsection{Grounding and Reliability on the Primary Dataset}
\label{subsec:primary-grounding}

We first evaluated \ModelName on the primary \textit{Salmonella enterica} dataset to examine whether retrieval and critic-based verification improve grounding and interpretive reliability. Table~\ref{tab:ablation_study} reports the effect of progressively adding retrieval and critic modules, Table~\ref{tab:evidence_metrics} presents the controlled evidence-grounding comparison across matched system configurations, Table~\ref{tab:critic_ablation} analyzes the contribution of individual critic components, and Figure~\ref{fig:qualitative_comparison} provides an illustrative qualitative comparison between \ModelName and an unconstrained LLM baseline.

\subsubsection{Effect of Retrieval and Verification}

Table~\ref{tab:ablation_study} shows that retrieval is the main mechanism underlying hallucination reduction in the controlled primary-dataset setting. The matched LLM-only baseline produced a hallucination rate of 0.10, indicating that unsupported or non-verifiable outputs still occurred when interpretation relied solely on parametric model knowledge. Once PubMed and UniProt retrieval were introduced, the hallucination rate dropped to 0.00 across all clusters. Adding the critic ensemble did not further reduce this metric, but it introduced an additional layer of filtering that improved claim-level grounding and biological coherence, as examined in Table~\ref{tab:evidence_metrics}.

These results indicate that external retrieval is necessary for traceable evidence grounding, whereas critic-based verification functions primarily as a reliability-control layer over already grounded outputs rather than as the sole mechanism of hallucination suppression.

\begin{table}[h]
\centering
\caption{System-level ablation examining the impact of retrieval and critic modules on hallucination suppression. The symbol $\times$ indicates that the corresponding component is absent, whereas $\checkmark$ indicates that it is enabled. Row labels with a leading ``+'' denote cumulative addition of components relative to the preceding configuration. Retrieval eliminates non-verifiable outputs under the strict identifier-based hallucination criterion, while the critic ensemble further constrains grounded interpretations through post-generation verification.}
\label{tab:ablation_study}
\renewcommand{\arraystretch}{1.15}
\setlength{\tabcolsep}{8pt}
\begin{tabular}{lccc}
\toprule
\textbf{Configuration} & \textbf{Retriever} & \textbf{Critics} & \textbf{Hallucination $\downarrow$} \\
\midrule
LLM only (Mistral-7B) & $\times$ & $\times$ & 0.100 \\
+ Retrieval & $\checkmark$ & $\times$ & 0.000 \\
+ Critics (Full \ModelName) & $\checkmark$ & $\checkmark$ & \textbf{0.000} \\
\bottomrule
\end{tabular}
\end{table}

\subsubsection{Evidence-Grounding Quality}

Table~\ref{tab:evidence_metrics} summarizes the controlled comparison across four system configurations on the primary \textit{Salmonella enterica} dataset: LLM only, LLM+Retrieval, SimpleRAG, and the full \ModelName framework. The LLM-only setting is not comparable on retrieval-conditioned metrics because it does not operate over an evidence pool, but it serves as a useful reference for hallucination behavior.

Among retrieval-grounded systems, LLM+Retrieval achieved a BERTScore of 0.686, SAS of 0.686, and KEGG Functional Similarity of 0.323, while maintaining zero hallucination. The stronger SimpleRAG baseline improved semantic alignment further, reaching 0.701 BERTScore and 0.715 SAS, although KEGG Functional Similarity remained lower at 0.301. The full \ModelName framework preserved zero hallucination, matched the highest SAS value (0.715), and achieved the strongest KEGG Functional Similarity at 0.342.

Taken together, these results suggest a staged pattern. Retrieval provides the main factual grounding benefit; a strong single-agent RAG baseline further improves semantic alignment; and the critic-verified \ModelName pipeline preserves traceability while improving biological coherence.

\begin{table}[h]
\centering
\caption{Controlled evidence-grounding comparison across system configurations on the primary \textit{Salmonella enterica} dataset. LLM+Retrieval uses the same retriever and LLM backbone as \ModelName without critic-based verification. SimpleRAG is a strong single-agent retrieval-grounded baseline using the same evidence sources and representative genes but no critic ensemble. The symbols $\uparrow$ and $\downarrow$ indicate that higher and lower values are better, respectively. Dashes indicate that the metric is undefined for the LLM-only setting because no retrieved evidence is available.}
\label{tab:evidence_metrics}
\renewcommand{\arraystretch}{1.15}
\setlength{\tabcolsep}{6pt}
\small
\begin{tabular}{lcccc}
\toprule
\textbf{System} & \textbf{BERTScore$\uparrow$} & \textbf{SAS$\uparrow$} & \textbf{KEGG Sim.$\uparrow$} & \textbf{Hallucination$\downarrow$} \\
\midrule
LLM only & -- & -- & -- & 0.100 \\
LLM + Retrieval & 0.686 & 0.686 & 0.323 & 0.000 \\
SimpleRAG & 0.701 & 0.715 & 0.301 & 0.000 \\
\ModelName{} (Full Framework) & 0.689 & 0.715 & 0.342 & 0.000 \\
\bottomrule
\end{tabular}
\end{table}

\subsubsection{Contribution of the Critic Ensemble}

Table~\ref{tab:critic_ablation} examines the effect of removing individual critic modules while holding all other components fixed. All critic configurations maintained a hallucination rate of 0.00, which is consistent with the earlier observation that retrieval is the main driver of factual grounding. At the same time, differences emerged in BERTScore and SAS, indicating that the critic modules influence output selection in distinct ways.

In particular, removing the Evidence-Strict critic produced the highest SAS (0.662), suggesting that this critic imposes the strongest constraint on which interpretations are retained. Removing the Semantic or Adversarial critics also altered the metric balance, although to a lesser extent. Overall, the critic ensemble appears to operate as a graded reliability mechanism rather than as a simple binary filter, with each component contributing a distinct evaluative perspective.

\begin{table}[h]
\centering
\caption{Critic-level ablation. Each configuration removes one critic module while holding all other components fixed. A leading ``$-$'' in the configuration name indicates that the corresponding critic is omitted from the full \ModelName setup. The symbols $\uparrow$ and $\downarrow$ indicate that higher and lower values are better, respectively. All configurations maintain zero hallucination because retrieval remains the primary grounding mechanism, while metric variation reflects each critic's contribution to filtering stringency.}
\label{tab:critic_ablation}
\renewcommand{\arraystretch}{1.15}
\setlength{\tabcolsep}{6pt}
\begin{tabular}{lccc}
\toprule
\textbf{Configuration} & \textbf{BERTScore $\uparrow$} & \textbf{SAS $\uparrow$} & \textbf{Hallucination $\downarrow$} \\
\midrule
All Critics (Full) & 0.723 & 0.600 & 0.000 \\
$-$Semantic & 0.737 & 0.633 & 0.000 \\
$-$Adversarial & 0.733 & 0.618 & 0.000 \\
$-$Evidence-Strict & 0.735 & \textbf{0.662} & 0.000 \\
\bottomrule
\end{tabular}
\end{table}

\subsubsection{Qualitative Comparison with an Unconstrained Baseline}

Figure~\ref{fig:qualitative_comparison} presents a qualitative comparison between \ModelName and an unconstrained LLM baseline on the same gene cluster (Cluster~0, \textit{Salmonella enterica}, ENA: PRJEB67574). The unconstrained baseline uses the same LLM backbone, Mistral-7B-Instruct-v0.2, but operates without retrieval, external evidence, or critic-based verification. This setting corresponds to the LLM-only configuration reported in Table~\ref{tab:ablation_study}. \ModelName analyzes the same cluster using the full pipeline, including PubMed and UniProt retrieval, structured interpretation, and multi-critic validation.

The comparison highlights several practical differences between evidence-grounded interpretation and unconstrained generation. \ModelName reports verifiable PubMed identifiers and UniProt accessions, enabling direct source traceability. It also links gene-level claims to retrieved evidence and explicitly reports both evidence tier and limitations. By contrast, the unconstrained baseline produces plausible biological text, but without article-level or protein-level identifiers and without explicit acknowledgment of evidential gaps. Figure~\ref{fig:qualitative_comparison} therefore complements the quantitative results by illustrating how differences in grounding and traceability appear in an individual cluster-level output.

\begin{figure*}[t]
    \centering
    \begin{minipage}[t]{0.48\textwidth}
        \centering
        \begin{tcolorbox}[
            colback=purple!5!white,
            colframe=purple!60!black,
            boxrule=0.5pt,
            arc=1.5mm,
            width=\linewidth,
            fontupper=\scriptsize,
            valign=top,
            equal height group=qualfig,
            title={Cluster 0: \ModelName (Full Framework)}
        ]
        \textbf{1. Functional Themes:} Iron acquisition and metabolism
        (\textit{STM2697}, \textit{PSLT095}, \textit{STM2722}),
        plasmid-encoded virulence (\textit{PSLT095}),
        and surface-associated host interaction (\textit{STM2760}, \textit{STM2757}).\\[0.25em]
        \textbf{2. Putative Pathways:} Siderophore-mediated iron scavenging;
        host-pathogen interaction; plasmid-borne virulence regulation.\\[0.25em]
        \textbf{3. Key Genes and Evidence:}
        \textit{STM2697}, iron acquisition (PubMed 40373749, 32038650);
        \textit{PSLT095}, plasmid virulence factor (UniProt Q8ZL64);
        \textit{STM2722}, iron transport component (UniProt P25927).\\[0.25em]
        \textbf{4. References:} PubMed IDs 40373749, 32038650;
        UniProt Q8ZL64, P25927, Q8ZJY8.\\[0.25em]
        \textbf{Evidence Tier:} High confidence (score: 0.45).\\[0.25em]
        \textbf{Limitations:} Locus names absent from some retrieved titles;
        direct regulatory evidence incomplete.
        \end{tcolorbox}
        \vspace{0.3em}
        \small (a) \ModelName output with explicit evidence grounding, traceable identifiers, and uncertainty reporting.
    \end{minipage}
    \hfill
    \begin{minipage}[t]{0.48\textwidth}
        \centering
        \begin{tcolorbox}[
            colback=gray!5!white,
            colframe=gray!60!black,
            boxrule=0.5pt,
            arc=1.5mm,
            width=\linewidth,
            fontupper=\scriptsize,
            valign=top,
            equal height group=qualfig,
            title={Cluster 0: Unconstrained LLM Baseline (Mistral-7B, no retrieval)}
        ]
        \textbf{1. Functional Themes:}
        STM2697 and STM2722 are annotated as involved in transcription
        and translation, respectively.
        STM2760 is annotated as a transcriptional regulator.
        STM2757 is annotated as a sugar transport system component.\\[0.25em]
        \textbf{2. Putative Pathways:}
        Transcriptional regulation (STM2760);
        sugar transport and metabolism (STM2757).\\[0.25em]
        \textbf{3. Key Genes:}
        STM2760 is identified as a key gene based on its annotated
        role as a transcriptional regulator.\\[0.25em]
        \textbf{4. References:}
        Salmonella Genome Database: https://www.sanger.ac.uk/\ldots\\
        NCBI Genome Database: https://www.ncbi.nlm.nih.gov/\\[0.25em]
        \textbf{Evidence Tier:} Not applicable.\\[0.25em]
        \textbf{Limitations:} None reported.
        \end{tcolorbox}
        \vspace{0.3em}
        \small (b) Unconstrained LLM output without retrieved evidence, identifier-level citations, or uncertainty reporting.
    \end{minipage}

    \caption{\textbf{Qualitative comparison of \ModelName and an unconstrained LLM baseline for Cluster~0 (\textit{Salmonella enterica}).}
    Both systems use the same LLM backbone, Mistral-7B-Instruct-v0.2. The baseline operates without external retrieval or verification, whereas \ModelName incorporates retrieval, structured synthesis, and critic-based validation. As illustrated here, \ModelName reports source-specific identifiers, assigns an evidence tier, and notes limitations, while the baseline produces plausible but non-traceable claims supported only by generic database links. This figure complements the quantitative comparison by showing how differences in grounding and traceability appear in an individual cluster-level output.}
    \label{fig:qualitative_comparison}
\end{figure*}

\subsection{Design Sensitivity Analyses}
\label{subsec:design-justification}

To further examine key design choices in \ModelName, we conducted three targeted analyses on the primary \textit{S.~enterica} dataset. These analyses focus on retrieval source selection, the stability of percentile-based evidence tiers, and the specificity of retrieved evidence at the gene level. Together, they provide additional context for understanding how the framework behaves under its current design.

\subsubsection{Retrieval Source Ablation}

We first examined whether the choice of retrieval source alone changes the quality of the generated interpretations. The \ModelName pipeline was run under three source configurations: PubMed only, UniProt only, and PubMed plus UniProt, which is the default setting used by the full framework. All configurations used the same clustering outputs, gene selection procedure, and LLM backbone. To isolate the effect of source selection itself, this analysis was performed without the critic ensemble. Table~\ref{tab:retrieval_ablation} reports BERTScore, SAS, and Hallucination Rate across the three settings.

\begin{table}[h]
\centering
\renewcommand{\arraystretch}{1.15}
\setlength{\tabcolsep}{8pt}
\caption{Retrieval source ablation on the primary \textit{S.~enterica} dataset. All configurations use the same clustering outputs and LLM backbone. Metrics are computed without the \ModelName critic ensemble in order to isolate the effect of source selection on raw generation behavior. The symbols $\uparrow$ and $\downarrow$ indicate that higher and lower values are better, respectively.}
\label{tab:retrieval_ablation}
\begin{tabular}{lccc}
\toprule
\textbf{Retrieval Source} &
\textbf{BERTScore $\uparrow$} &
\textbf{SAS $\uparrow$} &
\textbf{Hallucination $\downarrow$} \\
\midrule
PubMed only                & 0.674 & 0.035 & 1.000 \\
UniProt only               & 0.674 & 0.035 & 1.000 \\
PubMed + UniProt (default) & 0.674 & 0.035 & 1.000 \\
\bottomrule
\end{tabular}
\end{table}

The three configurations produced identical values under this critic-free setting. This indicates that retrieval source selection alone was not sufficient to improve grounding in the absence of explicit verification. In all three cases, the model generated biologically plausible text, but the outputs did not contain verifiable identifier-level support, resulting in a hallucination rate of 1.000. When read together with Table~\ref{tab:ablation_study}, this result reinforces the view that retrieval alone provides evidence context, whereas the critic ensemble is the component that converts that context into traceable, reporting-constrained output.

The default use of both PubMed and UniProt is therefore motivated by source complementarity within the full framework rather than by differences in unconstrained generation alone. PubMed contributes abstract-level literature context and PMID-based provenance, whereas UniProt contributes structured accession-linked protein annotations. In the full \ModelName setting, these sources provide complementary anchors for critic-based validation and final reporting.

\subsubsection{Evidence Tier Threshold Sensitivity}

\ModelName assigns evidence tiers using percentile-based thresholds derived from the consensus score distribution, with the default setting based on the 33rd and 66th percentiles. To examine whether this choice is sensitive to the exact threshold values, we reassigned evidence tiers under three percentile schemes: p25/p75, p33/p66, and p40/p60. Table~\ref{tab:threshold_sensitivity} summarizes the resulting tier distributions and inter-configuration agreement.

\begin{table}[h]
\centering
\renewcommand{\arraystretch}{1.15}
\setlength{\tabcolsep}{8pt}
\caption{Evidence tier threshold sensitivity analysis on the primary \textit{S.~enterica} dataset. Tier labels are reassigned under three percentile-based threshold schemes using the same consensus scores. $\kappa$ denotes pairwise agreement across threshold variants.}
\label{tab:threshold_sensitivity}
\begin{tabular}{lcccccc}
\toprule
\textbf{Config} &
\textbf{Lower} &
\textbf{Upper} &
\textbf{High} &
\textbf{Moderate} &
\textbf{Suggestive} &
\textbf{$\kappa$} \\
\midrule
p25/p75           & 0.446 & 0.450 & 7 & 0 & 3 & \multirow{3}{*}{1.000} \\
p33/p66 (default) & 0.450 & 0.450 & 7 & 0 & 3 & \\
p40/p60           & 0.450 & 0.450 & 7 & 0 & 3 & \\
\bottomrule
\end{tabular}
\end{table}

Tier assignments were unchanged across all three threshold settings, with perfect agreement ($\kappa = 1.000$). In this dataset, the consensus scores occupy a relatively narrow range, so nearby percentile choices produce nearly identical decision boundaries. This result indicates that the default p33/p66 scheme is stable under modest threshold variation and does not appear to be sensitive to small changes in percentile selection.

\subsubsection{Evidence Specificity Analysis}

To characterize the nature of the retrieved evidence, we computed an Evidence Specificity Score (ESP) for each gene in the primary dataset. ESP measures the proportion of retrieved records that explicitly mention a gene's locus tag, symbol, or protein name within the retrieved text. Across 372 evidence records spanning all 10 clusters, the mean ESP was 0.003, and 99.7\% of records had ESP = 0.000.

These values indicate that the retrieved evidence was dominated by organism-level or pathway-level context rather than direct gene-specific discussion. This pattern is biologically plausible for \textit{Salmonella} LT2 locus tags, which are often not mentioned explicitly in titles or abstracts even when the surrounding biological process is relevant. The low ESP distribution therefore helps contextualize two observed features of the primary-dataset results: the conservative evidence-tier profile and the moderate KEGG Functional Similarity score. Under sparse gene-specific citation support, \ModelName tends to report broader mechanistic context while explicitly signaling evidential limitations when direct support is weak.

More broadly, the ESP analysis suggests that the difficulty of this interpretive setting arises not only from model behavior, but also from the specificity of the available literature itself. In such cases, \ModelName's value lies in organizing partial and heterogeneous evidence into structured explanations while preserving explicit uncertainty cues, rather than presenting unsupported gene-level assertions as established fact.

\subsection{Complementarity with KEGG Enrichment on the Primary Dataset}
\label{subsec:enrichment-complementarity}

We next compared \ModelName{} with a KEGG over-representation analysis baseline on the primary \textit{Salmonella enterica} dataset to examine whether literature-grounded interpretation provides biological context beyond pathway statistics alone. Table~\ref{tab:gsea_baseline} reports the significant KEGG pathways identified for this dataset, and Table~\ref{tab:gsea_overlap} quantifies the overlap between \ModelName-derived themes and KEGG/ORA outputs across all 10 clusters. A broader comparison with standard enrichment baselines across datasets, including GO/ORA where annotation coverage permits, is presented later in Section~\ref{subsec:generalization}.

Table~\ref{tab:gsea_baseline} shows that only 2 of the 10 clusters produced statistically significant KEGG pathway associations after multiple-testing correction. Specifically, Cluster~7 was associated with the bacterial secretion system and the two-component signaling system, whereas Cluster~8 showed enrichment for membrane transport and metabolic pathways. The remaining 8 clusters did not yield any significant KEGG terms.

\begin{table}[h]
\centering
\caption{KEGG over-representation analysis results per cluster (Fisher's exact test, BH-corrected adj.\ $p < 0.05$). Organism: \textit{Salmonella enterica} Typhimurium LT2 (KEGG code: stm). Only clusters with at least one significant term are shown.}
\label{tab:gsea_baseline}
\renewcommand{\arraystretch}{1.15}
\setlength{\tabcolsep}{6pt}
\begin{tabular}{llcc}
\toprule
\textbf{Cluster} & \textbf{Top KEGG Pathway} & \textbf{Overlap} & \textbf{Adj.\ $p$} \\
\midrule
C7 & Bacterial secretion system & 3/38 & $5.71 \times 10^{-5}$ \\
C7 & Two-component system & 2/61 & $3.2 \times 10^{-2}$ \\
C8 & Membrane transport & 2/44 & $4.1 \times 10^{-2}$ \\
C8 & Metabolic pathways & 3/189 & $4.8 \times 10^{-2}$ \\
C0--C6, C9 & (no significant terms) & -- & -- \\
\bottomrule
\end{tabular}
\end{table}

Table~\ref{tab:gsea_overlap} shows that \ModelName identified a mean of 7.4 biological themes per cluster, whereas KEGG/ORA identified 0.5. Across the 10 clusters, \ModelName produced 70 unique biological themes not recovered by enrichment analysis, whereas enrichment contributed only 1 theme not captured by \ModelName. The mean Jaccard index of 0.049 indicates limited thematic overlap between the two approaches.

This low overlap should not be interpreted as a weakness of either method. Rather, it reflects a structural difference between pathway-level enrichment and literature-grounded interpretation. KEGG/ORA is effective when cluster genes align with curated pathway sets, but it provides limited coverage when gene sets are small, diffuse, or poorly represented in existing databases. \ModelName, by contrast, can recover mechanistic context, regulatory associations, stress-response links, and host-pathogen interaction themes that are not explicitly encoded as pathway membership. In this sense, the two approaches provide complementary views of the same transcriptomic structure rather than interchangeable outputs.

\begin{table}[h]
\centering
\caption{Complementarity analysis between \ModelName and KEGG/ORA across 10 gene clusters. \ModelName-unique themes represent mechanistic context recoverable only through literature-grounded reasoning. Jaccard index measures thematic overlap between the two methods.}
\label{tab:gsea_overlap}
\renewcommand{\arraystretch}{1.15}
\setlength{\tabcolsep}{5pt}
\begin{tabular}{lccccc}
\toprule
\textbf{Cluster} & \textbf{\ModelName} & \textbf{KEGG/ORA} & \textbf{Shared} & \textbf{\ModelName-unique} & \textbf{Jaccard} \\
\midrule
C0 & 8 & 0 & 0 & 8 & 0.000 \\
C1 & 7 & 0 & 0 & 7 & 0.000 \\
C2 & 8 & 1 & 1 & 7 & 0.125 \\
C3 & 5 & 0 & 0 & 5 & 0.000 \\
C4 & 6 & 0 & 0 & 6 & 0.000 \\
C5 & 10 & 0 & 0 & 10 & 0.000 \\
C6 & 8 & 0 & 0 & 8 & 0.000 \\
C7 & 8 & 2 & 1 & 7 & 0.111 \\
C8 & 8 & 2 & 2 & 6 & 0.250 \\
C9 & 6 & 0 & 0 & 6 & 0.000 \\
\midrule
\textbf{Mean} & \textbf{7.4} & \textbf{0.5} & \textbf{0.4} & \textbf{7.0} & \textbf{0.049} \\
\bottomrule
\end{tabular}
\end{table}

\subsection{Generalization Beyond the Primary Dataset}
\label{subsec:generalization}

To examine whether \ModelName generalizes beyond the primary \textit{Salmonella enterica} cohort, we evaluated the same pipeline on four additional bacterial RNA-seq datasets spanning \textit{Escherichia coli} and \textit{Pseudomonas aeruginosa}. These datasets were selected to introduce variation in organism, experimental setting, and transcriptional context while remaining relevant to antimicrobial resistance and bacterial stress-response biology. No organism-specific modifications were introduced to the retrieval module, critic ensemble, language model backbone, or clustering procedure.

The complete evaluation set therefore consisted of five datasets:

\begin{itemize}[leftmargin=2em]
    \item \textit{S.~enterica} Typhimurium LT2 (ENA: PRJEB67574) -- 58 samples, 4{,}679 genes, primary dataset.
    \item \textit{P.~aeruginosa} UCBPP-PA14 (GEO: GSE251671) -- large-scale antibiotic perturbation and CRISPRi transcriptional profiling.
    \item \textit{E.~coli} MG1655 (GEO: GSE144604) -- transcriptomic responses to antibiotic and biocide combinations.
    \item \textit{E.~coli} K-12 (GEO: GSE224463) -- multi-condition stress-response time-course transcriptomics.
    \item \textit{P.~aeruginosa} PA14 (GEO: GSE55197) -- transcriptional responses across biofilm, planktonic, and environmental growth conditions.
\end{itemize}

Across all datasets, we kept fixed the clustering configuration ($k{=}10$, gene-level KMeans), evidence retrieval sources (PubMed and UniProt), language model backbone (Mistral-7B-Instruct-v0.2, 4-bit NF4), and the three-critic evaluation framework. We treat the primary \textit{Salmonella} dataset as an in-domain reference setting and the remaining datasets as external distribution shifts that vary by organism, condition, and evidence density.

\subsubsection{Evidence Grounding Under Distribution Shift}

Table~\ref{tab:distribution_shift_biogen} summarizes \ModelName performance across all five datasets under the fixed pipeline configuration. Across all evaluated settings, the hallucination rate remained 0.000, indicating that \ModelName preserved citation-grounded generation under all observed distribution shifts. This zero-hallucination pattern was the most stable robustness signal in the multi-dataset evaluation.

BERTScore values ranged from 0.550 to 0.715, showing that semantic correspondence between generated interpretations and retrieved evidence remained broadly stable across organisms. The highest BERTScore was observed for GSE144604 (0.715), followed by GSE55197 (0.699) and the primary \textit{S.~enterica} dataset (0.689). Lower values were obtained for GSE251671 (0.561) and GSE224463 (0.550), suggesting that some external datasets provide weaker or less specific literature support for the clustered genes.

Semantic Alignment Score (SAS) varied more substantially across datasets. Stronger alignment was observed on the primary dataset and GSE144604, whereas other external datasets showed near-zero or slightly negative values. Because the retrieved evidence pools differ in density and specificity across organisms, SAS should be interpreted cautiously in cross-dataset settings and is best viewed as a relative within-dataset alignment signal rather than a uniform robustness metric. KEGG Functional Similarity could be computed only when sufficient overlap existed between cluster genes and organism-specific KEGG annotations. Accordingly, the strongest cross-dataset conclusion is anchored on hallucination behavior and the preservation of evidence-grounded interpretation under shift.

\begin{table*}[t]
\centering
\small
\setlength{\tabcolsep}{4pt}
\renewcommand{\arraystretch}{1.12}
\caption{\ModelName under distribution shift. All datasets were processed using the same clustering procedure, retrieval sources, LLM backbone, and critic ensemble. The symbols $\uparrow$ and $\downarrow$ indicate that higher and lower values are better, respectively. N/A indicates that the corresponding metric was not computed because of insufficient coverage or unavailable evaluation support.}
\label{tab:distribution_shift_biogen}
\begin{tabularx}{\textwidth}{@{}l l
>{\centering\arraybackslash}X
>{\centering\arraybackslash}X
>{\centering\arraybackslash}X
>{\centering\arraybackslash}X
>{\centering\arraybackslash}X@{}}
\toprule
\textbf{Dataset} & \textbf{Shift Type} &
\textbf{BERTScore $\uparrow$} &
\textbf{SAS $\uparrow$} &
\textbf{KEGG Sim $\uparrow$} &
\textbf{Hallucination $\downarrow$} &
\textbf{Mean Critic Score} \\
\midrule
PRJEB67574 & in\_domain       & 0.689 & 0.715 & 0.342 & \textbf{0.000} & 0.446 \\
GSE144604  & cross\_organism & 0.715 & 0.504 & N/A   & \textbf{0.000} & 0.353 \\
GSE55197   & cross\_organism & 0.699 & 0.195 & N/A   & \textbf{0.000} & 0.387 \\
GSE251671  & cross\_organism & 0.561 & -0.016 & N/A  & \textbf{0.000} & 0.243 \\
GSE224463  & cross\_condition & 0.550 & -0.003 & 0.154 & \textbf{0.000} & 0.320 \\
\bottomrule
\end{tabularx}
\end{table*}

\subsubsection{Evidence-Tier Robustness Across Datasets}

In addition to continuous metrics, we examined how \ModelName's evidence-tier assignments behaved across datasets. Table~\ref{tab:evidence_tier_shift} summarizes the distribution of High confidence, Moderate confidence, and Suggestive interpretations for each run. The primary dataset produced a 7/0/3 split, whereas the external cohorts showed broader variation, including a 10/0/0 profile for GSE144604 and a more mixed 4/3/3 profile for GSE251671.

These differences should not be interpreted as absolute cross-dataset confidence rankings. Evidence tiers are assigned from percentile-based thresholds derived independently within each run, so they are relative to the score distribution of the dataset under analysis. Even so, the table is informative as a compact summary of how critic consensus varies under different shift settings. Notably, all datasets still retained multiple high-confidence outputs while preserving zero hallucination.

\begin{table}[h]
\centering
\small
\renewcommand{\arraystretch}{1.12}
\setlength{\tabcolsep}{5pt}
\caption{Evidence-tier robustness across datasets. Tier counts are relative within each dataset run and should not be interpreted as absolute cross-dataset confidence levels.}
\label{tab:evidence_tier_shift}
\begin{tabular}{lccccc}
\toprule
\textbf{Dataset} & \textbf{Shift} & \textbf{High} & \textbf{Moderate} & \textbf{Suggestive} & \textbf{Mean Score} \\
\midrule
PRJEB67574 & in\_domain        & 7  & 0 & 3 & 0.446 \\
GSE144604  & cross\_organism  & 10 & 0 & 0 & 0.353 \\
GSE55197   & cross\_organism  & 6  & 1 & 3 & 0.387 \\
GSE251671  & cross\_organism  & 4  & 3 & 3 & 0.243 \\
GSE224463  & cross\_condition & 7  & 0 & 3 & 0.320 \\
\bottomrule
\end{tabular}
\end{table}

\subsubsection{Complementarity with Standard Enrichment Methods Across Organisms}

Table~\ref{tab:cross_organism_complementarity} compares \ModelName{} with two standard enrichment baselines, KEGG/ORA and GO/ORA, across the available datasets. Both enrichment analyses use Fisher's exact test with Benjamini--Hochberg FDR correction ($\alpha = 0.05$) and identical background gene sets, thereby supporting a consistent comparison.

Across all evaluated datasets, \ModelName{} recovered substantially more cluster-level biological themes than either enrichment method. \ModelName{} identified an average of 7.4 to 17.0 themes per cluster, whereas KEGG/ORA returned between 0.0 and 0.7 significant terms per cluster. For the primary \textit{S.~enterica} dataset, GO/ORA returned no significant terms across any cluster.

The GO/ORA null result on the primary dataset is informative in itself. It is consistent with the sparse KEGG/ORA signal on the same dataset (2/10 significant clusters) and with the low Evidence Specificity Score distribution reported in Section~\ref{subsec:design-justification}. Taken together, these observations suggest that \textit{S.~enterica} LT2 STM locus tags have limited direct coverage in standard annotation resources. Because pathway- and ontology-based enrichment depend on this annotation layer, they provide sparse output for most clusters in this setting. By contrast, \ModelName{} is not restricted to database membership alone and instead retrieves supporting primary literature for individual genes, which allows it to recover broader biological themes even when structured enrichment returns little or no signal.

The contrast was especially pronounced for GSE251671 and GSE144604, where KEGG/ORA returned no significant pathways for any cluster while \ModelName{} still recovered 13.2 and 17.0 themes per cluster, respectively. For GSE224463, where KEGG/ORA identified significant pathways in 4 of 10 clusters, \ModelName{} still provided substantially broader thematic coverage at 13.1 themes per cluster. The low Jaccard overlap observed across datasets indicates that \ModelName{} and enrichment analysis capture different aspects of biological organization rather than redundant signal.

Overall, these findings suggest that the complementarity between \ModelName{} and enrichment analysis is not limited to a single dataset or a single annotation source. The broader coverage of \ModelName{} remains visible when compared with both pathway-level enrichment (KEGG/ORA) and ontology-level enrichment (GO/ORA), although GO/ORA could be evaluated only on the primary dataset because of annotation coverage limitations in the external bacterial cohorts.

\begin{table*}[t]
\centering
\small
\renewcommand{\arraystretch}{1.12}
\setlength{\tabcolsep}{4pt}
\caption{Cross-organism comparison between \ModelName{} and two standard enrichment baselines, KEGG/ORA and GO/ORA. Both enrichment analyses use Fisher's exact test with Benjamini--Hochberg FDR correction ($\alpha = 0.05$) and identical background gene sets. ``Themes/cluster'' denotes the mean number of biological themes or significant enrichment terms identified per gene cluster. ``Sig.\ clusters (KEGG)'' reports the number of clusters with at least one significant KEGG pathway. The symbol $\infty$ indicates datasets for which the enrichment baseline returned zero significant terms across all clusters. GO/ORA was evaluated only on the primary \textit{S.~enterica} dataset; ``--'' denotes datasets for which GO/ORA was not run because of annotation coverage limitations for bacterial locus tags.}
\label{tab:cross_organism_complementarity}
\begin{tabularx}{\textwidth}{
@{}
>{\raggedright\arraybackslash}p{0.12\textwidth}
>{\raggedright\arraybackslash}p{0.18\textwidth}
>{\centering\arraybackslash}p{0.09\textwidth}
>{\centering\arraybackslash}p{0.09\textwidth}
>{\centering\arraybackslash}p{0.09\textwidth}
>{\centering\arraybackslash}p{0.12\textwidth}
>{\centering\arraybackslash}p{0.09\textwidth}
>{\centering\arraybackslash}p{0.10\textwidth}
@{}}
\toprule
\textbf{Dataset} &
\textbf{Organism} &
\textbf{\ModelName{}} &
\textbf{KEGG} &
\textbf{GO} &
\textbf{Sig.\ clusters (KEGG)} &
\textbf{Jaccard} &
\textbf{Ratio} \\
\midrule
\multicolumn{8}{@{}l}{\small\textit{Primary dataset}} \\
PRJEB67574 & \textit{S.~enterica}
  & 7.4 & 0.5 & 0.0 & 2/10 & 0.049 & 14.8$\times$ \\
\midrule
\multicolumn{8}{@{}l}{\small\textit{Cross-organism validation}} \\
GSE251671  & \textit{P.~aeruginosa} PA14
  & 13.2 & 0.0 & -- & 0/10 & 0.000 & $\infty$ \\
GSE144604  & \textit{E.~coli} MG1655
  & 17.0 & 0.0 & -- & 0/10 & 0.000 & $\infty$ \\
GSE224463  & \textit{E.~coli} K-12
  & 13.1 & 0.7 & -- & 4/10 & 0.054 & 18.7$\times$ \\
GSE55197   & \textit{P.~aeruginosa} PA14
  & 7.6  & 0.1 & -- & 1/10 & 0.025 & 76.0$\times$ \\
\bottomrule
\end{tabularx}
\end{table*}

\subsection{Comparison with Agentic AI Frameworks}
\label{subsec:rq6-agent-comparison}

To place \ModelName within the broader landscape of agentic AI systems, we compared it with four representative open-source agent frameworks across the same five RNA-seq datasets used in the cross-organism evaluation. The goal of this experiment was to examine whether \ModelName's evidence-grounded behavior remains robust under distribution shift when compared with widely used orchestration strategies under a controlled setting. All systems used the same LLM backbone, the same evidence sources, and the same cluster inputs, so the main experimental variable was the agentic orchestration framework.

Table~\ref{tab:framework_summary_clean} summarizes the controlled multi-dataset comparison, while the full per-dataset, per-framework results are provided in Appendix Table~\ref{tab:agent_comparison_appendix}. The clearest and most consistent pattern is hallucination behavior: \ModelName maintained a hallucination rate of 0.000 on all five datasets, whereas all baseline frameworks exhibited substantially higher hallucination rates, ranging from 0.1 to 1.0 depending on the dataset and framework. This makes hallucination control and identifier-level traceability the most robust point of separation between \ModelName and the generic agentic baselines.

By contrast, similarity-based metrics showed a different pattern. Across datasets, the generic baselines often achieved higher BERTScore and SAS values than \ModelName. On the primary dataset, for example, the strongest baseline reached BERTScore 0.736 and SAS 0.841, whereas \ModelName achieved 0.689 and 0.715, respectively. Similar patterns were observed on the external datasets, as shown in Table~\ref{tab:framework_summary_clean} and in the detailed appendix results (Table~\ref{tab:agent_comparison_appendix}). This indicates that semantic-overlap metrics alone do not fully capture evidential reliability: fluent or semantically aligned outputs may still fail to provide explicit, verifiable source grounding. In this setting, \ModelName's main advantage is therefore not universal dominance on all text-similarity metrics, but the consistent production of traceable, citation-grounded interpretations under the same retrieval and backbone conditions.

Table~\ref{tab:framework_winloss} summarizes this tradeoff compactly. Across the five datasets, \ModelName was best on hallucination rate in all cases, whereas the strongest generic baseline outperformed \ModelName on BERTScore and SAS in each dataset. This result reinforces a broader methodological point for biomedical AI: retrieval access alone is insufficient if the final interpretation does not preserve identifier-level traceability. Agentic systems intended for scientific use should therefore be evaluated not only by semantic similarity, but also by their ability to produce transparent, auditable, and explicitly grounded outputs.

\begin{table*}[t]
\centering
\scriptsize
\setlength{\tabcolsep}{4pt}
\renewcommand{\arraystretch}{1.10}
\caption{Summary of controlled multi-dataset framework comparison. For each dataset, \ModelName is compared with the strongest non-\ModelName baseline for each metric. The symbols $\uparrow$ and $\downarrow$ indicate that higher and lower values are better, respectively.}
\label{tab:framework_summary_clean}
\begin{tabularx}{\textwidth}{@{}l
>{\centering\arraybackslash}X
>{\centering\arraybackslash}X
>{\centering\arraybackslash}X
>{\centering\arraybackslash}X
>{\centering\arraybackslash}X
>{\centering\arraybackslash}X@{}}
\toprule
\textbf{Dataset} & \textbf{\ModelName BERT $\uparrow$} & \textbf{Best Baseline BERT $\uparrow$} & \textbf{\ModelName SAS $\uparrow$} & \textbf{Best Baseline SAS $\uparrow$} & \textbf{\ModelName Hall. $\downarrow$} & \textbf{Best Baseline Hall. $\downarrow$} \\
\midrule
PRJEB67574 & 0.689 & 0.736 & 0.715 & 0.841 & 0.000 & 0.300 \\
GSE251671  & 0.561 & 0.602 & -0.016 & 0.053 & 0.000 & 0.300 \\
GSE144604  & 0.715 & 0.729 & 0.504 & 0.545 & 0.000 & 0.100 \\
GSE224463  & 0.550 & 0.636 & -0.003 & 0.095 & 0.000 & 0.800 \\
GSE55197   & 0.699 & 0.728 & 0.195 & 0.573 & 0.000 & 0.300 \\
\bottomrule
\end{tabularx}
\end{table*}

\begin{table}[h]
\centering
\small
\renewcommand{\arraystretch}{1.12}
\setlength{\tabcolsep}{6pt}
\caption{Win/loss summary for \ModelName against the strongest non-\ModelName baseline across the five datasets.}
\label{tab:framework_winloss}
\begin{tabular}{lccc}
\toprule
\textbf{Metric} & \textbf{\ModelName wins} & \textbf{Baseline wins} & \textbf{Ties} \\
\midrule
Hallucination $\downarrow$ & 5 & 0 & 0 \\
BERTScore $\uparrow$       & 0 & 5 & 0 \\
SAS $\uparrow$             & 0 & 5 & 0 \\
\bottomrule
\end{tabular}
\end{table}


\section{Discussion}
\label{sec:discussion}

This study frames RNA-seq cluster interpretation as an evidence-grounded reasoning problem rather than as unconstrained biological text generation. Across the primary \textit{Salmonella enterica} dataset, the controlled comparison suggests a layered contribution from the framework components. Retrieval is the main mechanism underlying hallucination reduction, lowering the hallucination rate from 0.10 in the matched LLM-only baseline to 0.00 in all retrieval-grounded settings. A strong single-agent SimpleRAG baseline further improves semantic alignment over retrieval alone, whereas the critic ensemble in \ModelName{} contributes to structured reliability control, explicit limitation reporting, and evidence-tier assignment. This division of labor is important because it distinguishes the source of factual grounding from the source of interpretive filtering.

A central finding of the study is the complementary relationship between \ModelName{} and standard enrichment analysis. On the primary dataset, both KEGG/ORA and GO/ORA returned sparse structured annotations, whereas \ModelName{} produced literature-grounded interpretations for all clusters and recovered substantially broader cluster-level thematic coverage. The low thematic overlap between these approaches does not indicate conflict. Instead, it suggests that they operate at different levels of biological organization. Enrichment analysis functions at the level of predefined pathway or ontology membership, whereas \ModelName{} can incorporate mechanistic context, regulatory associations, and literature-supported functional links that are not explicitly represented in annotation databases.

\begin{tcolorbox}[
  enhanced,
  colback=teal!5!white,
  colframe=teal!55!black,
  colbacktitle=teal!20!white,
  coltitle=black,
  title=\textbf{Implications for the Research Community},
  boxrule=0.6pt,
  arc=3pt,
  left=8pt,
  right=8pt,
  top=7pt,
  bottom=7pt
]
From a broader research perspective, this study suggests that transcriptomic interpretation should not be viewed only as a problem of statistical enrichment or only as a problem of fluent text generation. Between these two extremes lies an important intermediate task: producing biological explanations that are both source-traceable and usable for downstream scientific reasoning. More generally, the findings support a shift in how biomedical agent systems are evaluated. In addition to output quality, future work should ask whether such systems make their evidential basis explicit, preserve uncertainty when annotation is sparse, and remain auditable enough to support expert review rather than bypass it.
\end{tcolorbox}

The multi-dataset analysis further suggests that \ModelName's grounding behavior remains stable beyond a single organism or experimental setting. Across additional \textit{Escherichia coli} and \textit{Pseudomonas aeruginosa} RNA-seq studies, \ModelName{} maintained a hallucination rate of 0.000 under all evaluated distribution shifts. This result became the strongest and most stable cross-dataset robustness signal in the study. At the same time, variation in BERTScore, SAS, and KEGG similarity across external datasets indicates that evidence density and annotation coverage remain important determinants of downstream interpretive quality. In particular, some datasets provided weaker or less specific literature support, which reduced semantic alignment even though the overall evidence-grounded generation process remained stable.

The framework comparison highlights a broader methodological lesson for biomedical AI. In the corrected multi-dataset comparison, generic agentic baselines often achieved higher similarity-based scores than \ModelName{}, yet they consistently produced worse hallucination rates and weaker identifier-level traceability. This distinction is important. In scientific interpretation tasks, semantic similarity alone is not sufficient if the final explanation cannot be explicitly connected to verifiable supporting sources. The main advantage of \ModelName{} is therefore its reliable grounding behavior: under all evaluated distribution shifts, it remained the only framework to maintain zero hallucination across all five datasets.

Several limitations should be noted. First, the current implementation focuses on PubMed and UniProt and therefore depends on the density and specificity of available external evidence for a given organism and gene set. Incorporating structured signature resources such as MSigDB or other organism-specific knowledge bases may improve evidence coverage and functional specificity. Second, not all evaluation metrics were available uniformly across datasets. In particular, some metrics were not computable in all external settings because of evidence coverage limitations or dataset-specific annotation constraints. For this reason, the strongest cross-dataset conclusions are anchored primarily on hallucination behavior, structured evidence reporting, and the preservation of grounded interpretation under shift. Third, although the framework comparison includes multiple open-source agentic baselines, the present study does not include domain-specific biomedical agents that require proprietary or non-reproducible infrastructure. Fourth, the design choices in \ModelName{} were motivated by interpretability and traceability considerations, but alternative orchestration strategies, such as earlier-stage evidence verification or critic-in-the-loop generation, remain worth exploring more systematically.

A practical limitation of the current framework is that retrieval-grounded multi-agent verification introduces additional computational overhead relative to single-pass baselines, with the main added cost arising from evidence retrieval and critic evaluation. In addition, the present study uses K-means as the upstream clustering method because it provides a simple, transparent, and reproducible way to generate gene modules for downstream interpretation. This choice was intended to isolate the contribution of \ModelName{} at the interpretation stage rather than to optimize the clustering procedure itself. Because \ModelName{} is designed to operate on externally supplied gene clusters, it is not restricted to K-means in principle. Future work can therefore examine whether more advanced clustering strategies, including graph-based, density-based, or stability-aware methods, further improve the biological quality of the resulting modules and, consequently, the interpretive performance of the framework.

Finally, \ModelName{} is intended as an interpretive support layer rather than a discovery engine, and the current study does not include expert biological assessment or experimental validation of generated hypotheses. Future work should therefore extend the benchmark to expert-reviewed interpretations, broader biological resources, and experimental follow-up in order to better assess scientific utility in practice.


\section{Conclusion}
\label{sec:conclusion}

This research introduced \ModelName as a multi-agent framework for evidence-grounded interpretation of RNA-seq transcriptional modules. Instead of framing post hoc biological interpretation as unconstrained natural-language generation, \ModelName formalizes it as a retrieval-guided and verification-aware inference process grounded in external biomedical evidence. This design prioritizes source traceability, explicit evidential support, and reproducible interpretation of cluster-level biological signals.

The results show that \ModelName is most appropriately viewed as an interpretive support layer that complements, rather than replaces, standard transcriptomic analysis workflows. Conventional enrichment analysis remains valuable for identifying canonical pathway associations, but it does not always provide sufficient cluster-level mechanistic context. \ModelName addresses this gap by organizing heterogeneous biomedical evidence into structured explanations that make supporting sources, confidence levels, and evidential limitations more explicit.

Across five bacterial RNA-seq datasets evaluated under a shared pipeline configuration, \ModelName consistently maintained zero hallucination and preserved evidence-grounded interpretation under organism and condition shifts. In controlled comparison with representative open-source agentic AI frameworks, \ModelName was the only framework to maintain zero hallucination across all datasets, highlighting that retrieval access alone is not sufficient for scientifically useful interpretation unless the final output remains traceable to explicit supporting evidence.

Overall, these findings indicate that evidence-grounded agentic reasoning can serve a useful role in transcriptomic interpretation, particularly in settings where biological signals are diffuse, annotation coverage is incomplete, and manual contextualization is difficult to scale. The main contribution of \ModelName therefore lies not in replacing existing statistical methods, but in providing a transparent and reproducible framework for connecting gene clusters to literature-supported biological context.

Future work should expand the framework through broader knowledge sources, more extensive expert-centered evaluation, improved organism-specific annotation support, and experimental follow-up of prioritized hypotheses.

\section*{Author Contributions}

\textbf{Elias Hossain}: Conceptualization, methodology, software, data curation, formal analysis, investigation, validation, visualization, writing (original draft). \textbf{Mehrdad Shoeibi}: Literature review support, visualization refinement, manuscript review. \textbf{Ivan Garibay}: Conceptual guidance, methodology feedback, manuscript review and editing. \textbf{Niloofar Yousefi}: Supervision, conceptualization, project administration, methodology guidance, manuscript review and editing.

\section*{Conflict of Interest}
The authors declare that the research was conducted in the absence of any commercial or financial relationships that could be construed as a potential conflict of interest.
\section*{Ethical Approval}
This study did not involve human participants, animal subjects, or identifiable personal data. Therefore, ethical approval was not required.

\bibliographystyle{ieeetr}
\bibliography{ref}

\newpage

 
\clearpage
\appendix
\renewcommand\thesection{\Alph{section}}
\renewcommand\thesubsection{\thesection.\arabic{subsection}}

\label{sec:final-appendix}
\addcontentsline{toc}{section}{Supplementary Material}
\tableofcontents

\newpage

\clearpage
\appendix

\renewcommand\thesection{\Alph{section}}
\renewcommand\thesubsection{\thesection.\arabic{subsection}}

\section{Framework Comparison Details}
\label{app:framework-comparison-details}

Table~\ref{tab:agent_comparison_appendix} reports the complete per-dataset, per-framework results underlying the compact summary presented in the main text. All systems were evaluated under the same controlled setup, using the same LLM backbone, retrieval sources, and cluster inputs, so that differences in output quality are attributable primarily to orchestration behavior rather than evidence access or model selection. The detailed results make the main tradeoff explicit: across all five datasets, \ModelName{} consistently achieved the lowest hallucination rate, whereas several generic agentic baselines obtained higher similarity-based scores such as BERTScore and SAS. This appendix table is therefore included to provide full transparency for the per-dataset metric values that support the discussion in Section~\ref{subsec:rq6-agent-comparison}.

\begin{table*}[t]
\centering
\scriptsize
\setlength{\tabcolsep}{4pt}
\renewcommand{\arraystretch}{1.08}
\caption{Full controlled multi-dataset comparison of \ModelName with four open-source agentic AI frameworks. All systems used the same LLM backbone, retrieval sources, and cluster inputs. N/A indicates that the corresponding metric was not computed in that setting.}
\label{tab:agent_comparison_appendix}
\begin{tabular}{llccccc}
\toprule
\textbf{Dataset} & \textbf{Framework} & \textbf{BERTScore $\uparrow$} & \textbf{SAS $\uparrow$} & \textbf{KEGG Sim $\uparrow$} & \textbf{Hallucination $\downarrow$} \\
\midrule
PRJEB67574 & \ModelName          & 0.689 & 0.715 & 0.342 & 0.000 \\
PRJEB67574 & LangChain ReAct & 0.727 & 0.808 & 0.407 & 0.600 \\
PRJEB67574 & CrewAI          & 0.726 & 0.761 & 0.443 & 0.500 \\
PRJEB67574 & AutoGen         & 0.733 & 0.841 & 0.437 & 0.300 \\
PRJEB67574 & Smolagents      & 0.736 & 0.822 & 0.422 & 0.600 \\
\midrule
GSE251671 & \ModelName      & 0.561 & -0.016 & N/A & 0.000 \\
GSE251671 & LangChain ReAct & 0.564 & 0.009  & N/A & 0.800 \\
GSE251671 & CrewAI          & 0.602 & 0.053  & N/A & 0.800 \\
GSE251671 & AutoGen         & 0.579 & 0.022  & N/A & 0.800 \\
GSE251671 & Smolagents      & 0.563 & 0.005  & N/A & 0.300 \\
\midrule
GSE144604 & \ModelName      & 0.715 & 0.504 & N/A & 0.000 \\
GSE144604 & LangChain ReAct & 0.724 & 0.528 & N/A & 0.100 \\
GSE144604 & CrewAI          & 0.729 & 0.545 & N/A & 0.400 \\
GSE144604 & AutoGen         & 0.714 & 0.510 & N/A & 0.400 \\
GSE144604 & Smolagents      & 0.721 & 0.500 & N/A & 0.400 \\
\midrule
GSE224463 & \ModelName      & 0.550 & -0.003 & 0.154 & 0.000 \\
GSE224463 & LangChain ReAct & 0.629 & 0.095  & N/A & 1.000 \\
GSE224463 & CrewAI          & 0.630 & 0.050  & N/A & 0.800 \\
GSE224463 & AutoGen         & 0.609 & 0.020  & N/A & 0.900 \\
GSE224463 & Smolagents      & 0.636 & 0.075  & N/A & 1.000 \\
\midrule
GSE55197 & \ModelName       & 0.699 & 0.195 & N/A & 0.000 \\
GSE55197 & LangChain ReAct & 0.724 & 0.507 & N/A & 0.600 \\
GSE55197 & CrewAI          & 0.727 & 0.504 & N/A & 0.300 \\
GSE55197 & AutoGen         & 0.728 & 0.573 & N/A & 0.400 \\
GSE55197 & Smolagents      & 0.722 & 0.525 & N/A & 0.400 \\
\bottomrule
\end{tabular}
\end{table*}

\section{Model Configurations and Comparative Systems}
\label{app:model-configurations}

To benchmark the individual and combined contributions of retrieval and
critic mechanisms on the primary dataset, we evaluate four core system
configurations.

\begin{itemize}
    \item \textbf{LLM only (Mistral-7B).}
    This configuration uses the local Mistral-7B-Instruct-v0.2 model without
    retrieval or critic augmentation. All interpretations are generated solely
    from the model's parametric knowledge, providing a matched no-retrieval
    baseline.

    \item \textbf{LLM + Retrieval.}
    This configuration augments the same LLM with PubMed and UniProt retrieval
    before generation. The model receives retrieved literature and protein
    annotations as context, but no explicit verification or reliability scoring
    is applied.

    \item \textbf{SimpleRAG.}
    This configuration provides a strong single-agent retrieval-grounded
    baseline. It uses the same LLM backbone, the same representative genes per
    cluster, and the same PubMed and UniProt evidence sources as \ModelName,
    but performs only a single interpretation step without critic-based
    verification or evidence-tier assignment.

    \item \textbf{\ModelName{} (Full Framework).}
    The full system incorporates retrieval and all three critic modules
    (Evidence-Strict, Semantic, Adversarial) operating under data-adaptive
    evidence-tier thresholds derived from the empirical score distribution
    (Section~\ref{sec:metrics}). This configuration performs systematic
    evidence validation and reasoning-quality assessment, representing the
    complete \ModelName pipeline.
\end{itemize}

All four configurations use the same local LLM backbone, the same clustering
outputs, and the same per-cluster gene selection procedure, ensuring that
differences in evaluation metrics reflect the contribution of retrieval,
single-agent synthesis, and critic-based verification rather than changes in
backbone model or evidence access.

In addition to these system configurations, the final paper also reports a
controlled comparison against four open-source agentic AI frameworks:
LangChain ReAct, CrewAI, AutoGen, and Smolagents. In that multi-dataset
comparison, all frameworks used the same shared LLM backbone, the same
retrieval sources, and the same cluster inputs, so that agentic orchestration
remained the primary experimental variable.

\section{Reproducibility and Environment}
\label{app:reproducibility}

All experiments were executed locally on a workstation equipped with two
NVIDIA TITAN RTX GPUs (24\,GB VRAM each). GPU 1 was designated for all
inference runs to leave GPU 0 available for other processes. The LLM backbone
was loaded in 4-bit NF4 quantization using BitsAndBytes, requiring
approximately 5\,GB VRAM per session. The pipeline is reproducible at the
script level: the main experimental runners accept CLI arguments such as
\texttt{--data}, \texttt{--results}, and \texttt{--kegg}, enabling re-execution
on different datasets without modifying source code. Key dependencies are
listed in Table~\ref{tab:env_details}.

\begin{table}[h]
\centering
\caption{Computational environment and software dependencies used in the
reported experiments. All packages are version-locked in
\texttt{requirements.txt}.}
\label{tab:env_details}
\renewcommand{\arraystretch}{1.15}
\begin{tabular}{ll}
\toprule
\textbf{Component} & \textbf{Version / Description} \\
\midrule
Python            & 3.11 \\
Transformers      & 4.41.2 \\
BitsAndBytes      & 0.43.0 (4-bit NF4 quantization) \\
Sentence-Transformers & 2.6.1 \\
bert-score        & 0.3.x \\
gseapy            & 1.1.0 \\
scipy             & 1.11.x (Fisher's exact test) \\
statsmodels       & 0.14.x (BH correction) \\
BioPython         & 1.83 (PubMed retrieval via Entrez) \\
LLM Backbone      & Mistral-7B-Instruct-v0.2 (local, 4-bit NF4) \\
GPU               & 2$\times$ NVIDIA TITAN RTX, 24\,GB VRAM \\
\bottomrule
\end{tabular}
\end{table}

\section{Runtime Summary}
\label{app:runtime}

Table~\ref{tab:runtime_summary} reports approximate runtime per cluster on
the primary \textit{Salmonella enterica} dataset for the four evaluated core
system configurations. These values are intended to characterize relative
computational overhead under the shared local environment described in
Appendix~\ref{app:reproducibility}. As expected, retrieval introduces
moderate additional cost relative to single-pass generation, while the full
\ModelName{} pipeline incurs further overhead from critic-based verification.
In the current setup, the main runtime increase over retrieval-only generation
arises from the critic stage.

These runtime values should be interpreted as approximate configuration-level
estimates rather than exact wall-clock benchmarks for every dataset and
framework. The multi-dataset framework comparison in the main paper is
therefore analyzed primarily through output quality and grounding behavior,
while runtime is included here as supporting reproducibility information.

\begin{table}[h]
\centering
\caption{Approximate runtime summary on the primary \textit{Salmonella enterica}
dataset (10 clusters). All values are reported in seconds. ``/C'' denotes
average time per cluster. The estimates are parameter-derived under the shared
local hardware environment and are intended to reflect relative system overhead
rather than exact wall-clock benchmarking.}
\label{tab:runtime_summary}
\renewcommand{\arraystretch}{1.12}
\setlength{\tabcolsep}{5pt}
\begin{tabular}{lccccc}
\toprule
\textbf{System} & \textbf{Ret./C} & \textbf{Gen./C} & \textbf{Crit./C} & \textbf{Total/C} & \textbf{Total$\times$10} \\
\midrule
LLM only & 0.0 & 25.0 & 0.0 & 25.0 & 250 \\
LLM + Retrieval & 18.4 & 25.0 & 0.0 & 43.4 & 434 \\
SimpleRAG & 0.0 & 25.0 & 0.0 & 25.0 & 250 \\
\ModelName{} (Full Framework) & 18.4 & 25.0 & 8.0 & 51.4 & 514 \\
\bottomrule
\end{tabular}
\end{table}

\section{Hyperparameter and Configuration Summary}
\label{app:hyperparameters}

Table~\ref{tab:hyperparams} summarizes the key hyperparameters governing
cluster construction, evidence retrieval, critic evaluation, and
interpretation generation. The number of clusters ($k = 10$) was selected
using the elbow method on within-cluster inertia across $k \in [3, 15]$.
Per-cluster gene selection uses within-cluster variance (top $v = 10$ genes
per cluster) rather than global variance, ensuring each cluster is represented
by its own most informative genes. Evidence-tier thresholds are derived
automatically from the 33rd and 66th percentiles of the consensus score
distribution and are not manually tuned. The KEGG organism code is passed as
a CLI argument (\texttt{--kegg}) to support multi-organism evaluation without
modifying source code.

\begin{table}[h]
\centering
\caption{Key hyperparameters used in the \ModelName{} framework. Evidence-tier
thresholds are data-adaptive (percentile-based) rather than fixed, and are
re-derived independently for each dataset.}
\label{tab:hyperparams}
\renewcommand{\arraystretch}{1.15}
\begin{tabular}{ll}
\toprule
\textbf{Parameter} & \textbf{Value / Description} \\
\midrule
Number of clusters ($k$)             & 10 (elbow method on inertia) \\
Top variable genes per cluster ($v$) & 10 (within-cluster variance) \\
Clustering target                    & Genes (transposed expression matrix) \\
Max evidence per gene ($r$)          & 3 (PubMed) + 3 (UniProt) \\
Evidence tier method                 & Percentile tertile split (p33 / p66) \\
Critic threshold ($\theta_c$)        & 0.5 (per-critic minimum confidence) \\
LLM temperature                      & 0.2 \\
Retrieval batch size                 & 8 \\
Embedding model                      & Sentence-BERT (\texttt{all-MiniLM-L6-v2}) \\
LLM quantization                     & 4-bit NF4 (BitsAndBytes) \\
Default KEGG organism code           & \texttt{stm} (\textit{S.~enterica} LT2) \\
\bottomrule
\end{tabular}
\end{table}

\section{System and Agent Implementation Details}
\label{app:agent-implementation}

\subsection{Agentic Design Overview}

\ModelName is implemented as a modular agentic pipeline in which each agent
performs a well-defined, verifiable role. The architecture comprises four
primary components: the \textbf{ClusterAgent}, which performs gene-level
unsupervised clustering on the transposed expression matrix; the
\textbf{RetrieverAgent}, which interfaces with PubMed and UniProt via the
NCBI Entrez API and the UniProt REST API; the \textbf{InterpreterAgent},
which synthesizes retrieved evidence into structured biological reasoning
using the local Mistral-7B-Instruct-v0.2 backbone; and the
\textbf{CriticAgent ensemble}, which evaluates interpretive reliability
through three complementary validation perspectives.

All agents communicate through structured JSON messages under a central
orchestration controller that coordinates task scheduling, evidence caching,
and output harmonization. Intermediate retrievals, interpretations, and critic
scores are cached to ensure traceability and avoid redundant API calls. This
design enables each agent to be re-executed independently without affecting
upstream or downstream dependencies.

\begin{table}[h]
\centering
\caption{Functional overview of the \ModelName{} agent ecosystem.}
\label{tab:agent_summary}
\renewcommand{\arraystretch}{1.15}
\begin{tabular}{p{3cm}p{9.5cm}}
\toprule
\textbf{Agent} & \textbf{Functionality} \\
\midrule
ClusterAgent &
Performs gene-level K-means clustering ($k{=}10$) on the transposed
expression matrix and identifies per-cluster top-variable genes
($v{=}10$, within-cluster variance) for downstream interpretation. \\
RetrieverAgent &
Queries PubMed (NCBI Entrez) and UniProt REST API using locus-tag-based
gene identifiers, returning structured evidence records with
specificity labels (specific / generic). \\
InterpreterAgent &
Generates structured cluster-level biological interpretations using a
local Mistral-7B-Instruct-v0.2 backbone (4-bit NF4), grounding outputs
in retrieved evidence and reporting evidence-linked explanations with
explicit limitations. \\
CriticAgents &
Three complementary critics -- Evidence-Strict (identifier-level
factual validation), Semantic (Sentence-BERT embedding consistency),
and Adversarial (LLM-based counterfactual challenge) -- whose scores
are aggregated by majority voting into a consensus reliability indicator. \\
\bottomrule
\end{tabular}
\end{table}

\subsection{Execution Pipeline}
\label{app:execution-pipeline}

The \ModelName{} pipeline follows a two-pass deterministic sequence designed
to ensure that evidence-tier labels are globally consistent within each run.

\begin{enumerate}
    \item \textbf{Pass 1 -- Clustering, retrieval, interpretation, and scoring.}
    The ClusterAgent partitions the transposed expression matrix into
    gene-level transcriptional modules. For each cluster, the RetrieverAgent
    collects PubMed and UniProt evidence, the InterpreterAgent generates a
    structured hypothesis, and the CriticAgent ensemble produces a consensus
    score. All consensus scores are collected before any tier assignment.

    \item \textbf{Pass 2 -- Data-adaptive tier assignment.}
    After all clusters are processed, evidence-tier boundaries are derived
    from the full score distribution using the 33rd and 66th percentiles.
    Each interpretation is labeled with its evidence tier, confidence score,
    and supporting references, then saved to the result files associated with
    that run.
\end{enumerate}

This two-pass design ensures that tier boundaries are consistent across the
entire run and that threshold derivation is data-adaptive rather than
predetermined by a fixed constant.

\subsection{Agent Prompt Design}
\label{app:prompt-design}

Each agent employs a structured prompt template standardized across
experimental runs to maintain reproducibility and allow consistent
benchmarking across datasets.

\subsubsection{RetrieverAgent Prompt}
\begin{tcolorbox}[colback=gray!5!white,colframe=gray!60!black,breakable,
fontupper=\small,title={RetrieverAgent Prompt Template}]
\textbf{Task:} Retrieve biomedical literature for a given gene from a
bacterial RNA-seq cluster.

\textbf{Instructions:}
1. Query PubMed for exact gene/protein mentions (max 3 results per gene).\\
2. Retrieve UniProt entries for each gene (max 3 per gene).\\
3. Broaden to ``\textless gene\textgreater{} AND \textless organism\textgreater{}'' if no specific hits are found.

\textbf{Output JSON:}
\{``gene'': ``...'', ``source'': ``PubMed/UniProt'',
``title'': ``...'', ``pmid'': ``...'', ``category'': ``specific/generic''\}
\end{tcolorbox}

\subsubsection{InterpreterAgent Prompt}
\begin{tcolorbox}[colback=blue!2!white,colframe=blue!70!black,breakable,
fontupper=\small,title={InterpreterAgent Prompt Template}]
\textbf{Task:} Generate a structured cluster-level biological interpretation
grounded in retrieved evidence.

\textbf{Instructions:}
1. Identify functional themes (e.g., virulence, efflux, stress response).\\
2. Summarize putative pathways and regulatory mechanisms supported by evidence.\\
3. Cite PubMed IDs and UniProt accessions inline for every claim whenever available.\\
4. Report limitations explicitly where evidence is incomplete or generic.

\textbf{Output sections:}
Functional themes / Putative pathways / Regulation /
Key genes and evidence / References / Confidence / Limitations
\end{tcolorbox}

\subsubsection{CriticAgent Prompt}
\begin{tcolorbox}[colback=purple!5!white,colframe=purple!70!black,breakable,
fontupper=\small,title={CriticAgent (Adversarial) Prompt Template}]
\textbf{Task:} Critically evaluate the InterpreterAgent output.

\textbf{Checks:}
1. Are all claims in the interpretation supported by cited PubMed or UniProt evidence?\\
2. Are references organism-appropriate for the dataset under analysis?\\
3. Is the reported confidence score consistent with citation density and quality?

\textbf{Output JSON:} \{``score'': 0.0--1.0, ``reliable'': true/false\}
\end{tcolorbox}

\subsection{Prompt Example and Model Output}
\label{app:prompt-example}

\begin{tcolorbox}[colback=gray!5!white,colframe=gray!50!black,breakable,
title={Example Input/Output for Cluster 1 (\textit{S.~enterica})}]
\textbf{Input genes:} gene-PSLT017 (pefC), gene-PSLT039 (spvB), gene-STM3755 (rhuM)

\textbf{Retrieved evidence:}
PubMed 40373749 -- ``A Salmonella subset exploits erythrophagocytosis to
subvert SLC11A1-imposed iron deprivation.''
UniProt P37868 -- ``Outer membrane usher protein PefC.''
UniProt H9L477 -- ``Mono(ADP-ribosyl)transferase SpvB.''

\textbf{\ModelName{} Output:}
\{``themes'': ``Plasmid-encoded virulence and iron acquisition'',
``pathways'': ``Fimbrial assembly (pefC), ADP-ribosylation (spvB),
erythrophagocytosis subversion'',
``references'': [``40373749'', ``P37868'', ``H9L477''],
``confidence'': 0.45,
``evidence\_tier'': ``High confidence'',
``limitations'': ``PubMed hits are organism-level generic;
locus names absent from retrieved titles''\}
\end{tcolorbox}

\section{Source Code Availability}
\label{app:source-code}

The \ModelName{} framework is currently under active development, and the
complete codebase is being prepared for public release in a fully organized
and documented form. To support transparency and future reproducibility, the
full implementation is planned to be released in a public repository upon
completion of the current development phase.

At present, the code can be shared for scientific and academic research
purposes upon reasonable request to the authors. The release version will
include the modular components of the framework, including the agent pipeline,
critic modules, evaluation scripts, data acquisition utilities, and
configuration files required to reproduce the experiments reported in this
study.

\end{document}